\documentclass{article}

\usepackage{arxiv}

\usepackage{graphicx}
\usepackage{cite}
\usepackage{amsmath,amssymb,amsfonts}
\usepackage{hyperref}
\usepackage{mathrsfs}

\newsavebox{\savepar}
\newenvironment{boxit}{\begin{center} \begin{lrbox}{\savepar}
			\begin{minipage}[b]{140mm}}
			{\end{minipage}\end{lrbox}
		\fbox{\usebox{\savepar}
} \end{center}}

\def\Z{{\sf Z}}
\def\G{{\sf G}}
\def\mod{{\sf mod}}

\title{On the Security of A Remote Cloud Storage Integrity Checking Protocol}

\author{Faen Zhang, Xinyu Fan, Pengcheng Zhou, Wenfeng Zhou\\
	\{faenzhang, pengchengzhou\}ainnovation@gmail.com\\
	fanxinyu@ainnovation.com, zhouwenfeng@interns.ainnovation.com\\
	AInnovation Technology Ltd.} 
\begin{document}
\maketitle

\begin{abstract}
	Data security and privacy is an important but challenging
	problem in cloud computing. One of the security concerns from
	cloud users is how to efficiently verify the integrity of their
	data stored on the cloud server. Third Party Auditing (TPA) is a
	new technique proposed in recent years to achieve this goal. In a recent paper (IEEE Transactions on Computers 62(2): 362-375
	(2013)), Wang {\em et al.} proposed a highly efficient and
	scalable TPA protocol and also a Zero Knowledge Public Auditing
	protocol which can prevent offline guessing attacks. However, in
	this paper, we point out several security weaknesses in Wang {\em et al.}'s protocols: first, we show that an attacker can
	arbitrarily modify the cloud data without being detected by the
	auditor in the integrity checking process, and the attacker can
	achieve this goal even without knowing the content of the cloud
	data or any verification metadata maintained by the cloud server; secondly, we show that the Zero Knowledge Public Auditing protocol cannot achieve its design goal, that is to prevent offline guessing attacks.
\end{abstract}
\section{Introduction}
Cloud computing offers different types of computational services
to end users via computer networks \cite{liu2015application}. It will become the next-generation information technology architecture due to its long list of advantages. Cloud storage\cite{hashem2015rise} is one of the major services in cloud computing where user data are stored and maintained by cloud servers. It allows users to access their data via computer networks at anytime and from anywhere.

Despite the great benefits provided by cloud computing\cite{sari2015review}, data security is a very important but challenging problem that must be solved \cite{zissis2012addressing}. One of the major concerns of data security is data integrity. Data integrity check can be done by periodically examining the data files stored on the cloud server if the data owner possesses some verification token (e.g. a hash digest of the data file). However, such an approach can be very expensive if the amount of data is huge. An interesting problem is to check data integrity remotely without the need of accessing the full copy of data stored on the cloud server.

Several techniques, such as Proof of Retrievability (POR)
\cite{ShachamW08,JuelsK07} and Third Party Auditing (TPA)
\cite{WangRLL10,WangWRLL11,WangChowWangRenLou}, have been proposed to solve the above data integrity checking problem. POR is loosely speaking a kind of Proof of Knowledge (POK) \cite{BellareG92} where the knowledge is the data file, while TPA allows any third party (or auditor) to perform the data integrity checking on behalf of the data owner just based on some public information (e.g. the data owner's public key).

In INFOCOM'10, Wang {\em et al.} \cite{WangWRL10} proposed a
privacy-preserving public auditing protocol with high efficiency and scalability. In particular, the proposed protocol supports batch auditing, which means the third party auditor can concurrently handle simultaneous auditing of multiple tasks.
In \cite{WangChowWangRenLou}, Wang {\em et al.} further extended
their TPA protocol and proposed a new Zero Knowledge Public
Auditing (ZKPA) protocol. The main security goal of the ZKPA
protocol is to prevent offline guessing attack (or offline
dictionary attack). It is worth noting that the early version of
Wang {\em et al.}'s TPA protocol published in INFOCOM'10 is
insecure: Xu {\em et al.} showed \cite{XuHeAb12} that the cloud
server can modify the user data without being caught by the
auditor in the auditing process. However, Xu {\em et al.}'s attack cannot be applied to the TPA and ZKPA protocols in
\cite{WangChowWangRenLou}.

In this paper, we show that there are several security weaknesses in Wang {\em et al.}'s TPA and ZKPA protocols
\cite{WangChowWangRenLou}. First, we show that an attacker can
arbitrarily modify the cloud data without being detected by the
auditor in the integrity checking process of both protocols. We
show that such an attack can be performed by different types of
attackers under different scenarios, and in the weakest attacking setting, the attack can be launched even when the attacker doesn't know the content of the cloud data or any verification metadata which are maintained by the cloud server and required in the auditing process (the only information the attacker needs to know is how data are modified). In reality, such an attack can be launched by either external or internal attacks (e.g. a malicious programmer who doesn't have access to cloud user data can perform such an attack by embedding some software bugs in a computer program on the cloud server). We remark that it is possible to prevent such attacks by using some extra security mechanisms (e.g. access control) on the cloud server, but this is orthogonal to the security goals of an integrity checking scheme. The key point is that, if an integrity checking protocol is secure and robust, then
once the user data stored on the cloud server have been modified, the auditor must be able to detect it in the integrity checking process. Secondly, we show that the ZKPA protocol cannot achieve its original security goal, that is, we can still launch an offline guessing attack against the protocol.

For the following part of this paper, we first review Wang et al.'s threat model and their TPA and ZKPA protocols. Then we show the security weaknesses in these two protocols in Sec.~\ref{sec:attack}.

\section{Review of Wang \emph{et al.}'s Threat Model and Protocols}
\subsection{The Threat Model}\label{sec:threatmodel}

We briefly review the threat model presented in
\cite{WangChowWangRenLou}. The cloud data storage service involves three entities: the cloud server, the cloud user, and the third party auditor (TPA). The cloud user relies on the cloud server to store and maintain his/her huge amount of data. Since the user no longer keeps the data locally, it is of critical importance for the user to ensure that the data are correctly stored and maintained by the cloud server. In order avoid periodically data integrity verification, the user may resort to a TPA for checking the integrity of his/her outsourced data. However, the data must be kept secret from the TPA during the integrity checking process.

In \cite{WangChowWangRenLou}, it is assumed that data integrity
threats can come from both internal and external attacks to the
cloud server, such as malicious software bugs, hackers, network
bugs, etc. Besides, the cloud server may also try to hide data
corruption incidents to users for the sake of reputation. However, it is assumed that the TPA is reliable and independent, and would not collude with the cloud server.

Five security goals are listed in \cite{WangChowWangRenLou}:
public auditability, storage correctness, privacy preserving,
batch auditing, and lightweight. Among these goals, storage
correctness and privacy preserving (i.e. the auditor cannot learn the content of the user data in the auditing process) are the most important security goals that must be achieved by a
privacy-preserving third party auditing protocol. For Zero
Knowledge Public Auditing, there is an extra security goal, that
is the protocol must be secure against offline guessing attacks.

\subsection{Notations and Preliminaries}

Before describing Wang \emph{et al.}'s TPA and ZKPA protocols, we
first introduce some notations and tools used by these protocols.

\begin{itemize}
	\item   $F$: the data file to be outsourced, denoted as a sequence of $n$ blocks $m_1, ..., m_n \in \Z_p $ for some large prime $p$.
	%\item $MAC$: Message Authentication Code (MAC) function, defined
	%as $MAC_{(\cdot)}(\cdot): {\cal K} \times \{0,1\}^* \rightarrow \{0,
	%1\}^l$ where ${\cal K}$ denotes the key space.
	\item $H(\cdot),h(\cdot)$: cryptographic hash functions.
\end{itemize}

\smallskip\noindent\textbf{Bilinear Map.} Let $\G_1, \G_2$ and $\G_T$ be
multiplicative cyclic groups of prime order $p$. Let $g_1$ and $g$
be generators of $\G_1$ and $\G_2$, respectively. A bilinear map is
a map $e: \G_1 \times \G_2 \rightarrow \G_T$ such that for all $u\in
\G_1$, $v \in \G_2$ and $a,b \in \Z_p$, $e(u^a,v^b) = e(u,v)^{ab}$.
Also, the map $e$ must be efficiently computable and non-degenerate
(i.e. $e(g_1, g) \ne 1$).

\subsection{The Third Party Auditing Protocol}

Let $(p, \G_1, \G_2, \G_T, e, g, H, h)$ be the system parameters as
introduced above. Wang \emph{et al.}'s privacy-preserving public
auditing scheme works as follows:
\medskip

\noindent {\bf Setup Phase}:

\noindent {KeyGen}: The cloud user runs KeyGen to generate the
public and secret keys. Specifically, the user generates a random
verification and signing key pair $(spk,ssk)$ of a digital signature
scheme, a random $x \leftarrow \Z_p$, a random element $u \leftarrow
\G_1$, and computes $v \leftarrow g^x$. The user secret key is $sk =
(x,ssk)$ and the user public key is $pk = (spk, v, g, u, e(u, v))$.
\medskip

\noindent {SigGen}: Given a data file $F = (m_1,...,m_n)$, the user
first chooses uniformly at random from $\Z_p$ a unique identifier
$name$ for $F$. The user then computes authenticator $\sigma_i$ for
each data block $m_i$ as $\sigma_i \leftarrow (H(W_i) \cdot
u^{m_i})^x \in \G_1$ where $W_i = name\|i$. Denote the set of
authenticators by $\phi = \{\sigma_i\}_{1 \leq i \leq n}$. Then the
user computes $t = name\|SSig_{ssk}(name)$ as the file tag for $F$,
where $SSig_{ssk}(name)$ is the user's signature on $name$ under the
signing key $ssk$. It was assumed that the TPA knows the number of
blocks $n$. The user then sends $F$ along with the verification
metadata $(\phi,t)$ to the cloud server and deletes them from local
storage.

\begin{figure*}[ht]
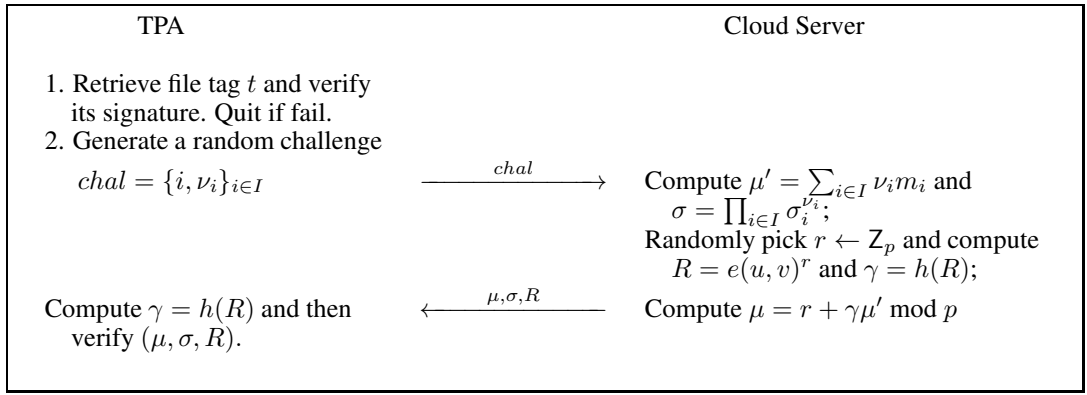

	\begin{boxit}
		\begin{center}
			\begin{tabular}{l@{\hspace{0.5cm}}c@{\hspace{0.5cm}}l}
				{ ~~~~~~~~~~~~~TPA}            &       &   { ~~~~~~~~~~~Cloud Server }\\
				%\hline
				&   &\\
				1. Retrieve file tag $t$ and verify            &       &  \\
				~~~~its signature. Quit if fail.   &   &  \\
				2. Generate a random challenge && \\
				~~~~ $chal = \{i, \nu_i\}_{i\in I}$  &  $\stackrel{chal}{\overrightarrow{\hspace{2.5cm}}}$          &  Compute $\mu' = \sum_{i\in I} \nu_i m_i$ and \\
				&& ~~~~$\sigma = \prod_{i\in I}\sigma_i^{\nu_i}$;  \\
				&& Randomly pick $r \leftarrow \Z_p$ and compute\\
				&& ~~~~$R=e(u,v)^r$ and $\gamma = h(R)$; \\
				Compute $\gamma=h(R)$ and then  &  $\stackrel{\mu, \sigma, R}{\overleftarrow{\hspace{2.5cm}}}$ &  Compute $\mu = r + \gamma\mu' \mbox{ mod } p$\\
				~~~~verify $(\mu, \sigma, R).$ &&\\
				& &\\
			\end{tabular}
		\end{center}
	\end{boxit}
	\caption{The third party auditing protocol by Wang {\em et al.}
		\cite{WangChowWangRenLou}.} \label{fig:wang}
\end{figure*}

\medskip
\noindent {\bf Audit Phase (Figure~\ref{fig:wang})}:

\noindent VerifySig: The TPA first retrieves the file tag $t$ and
verifies the signature $SSig_{ssk}(name)$ by using $spk$. The TPA
quits by emitting FALSE if the verification fails. Otherwise, the
TPA recovers $name$.
\medskip

\noindent Challenge: The TPA generates a challenge $chal$ for the
cloud server as follows: first pick a random $c$-element subset $I =
\{s_1,...,s_c\}$ of set $[1,n]$, and then for each element $i \in
I$, choose a random value $\nu_i$. The TPA sends $chal =
\{(i,\nu_i)\}_{i\in I}$ to the cloud server.

\medskip
\noindent GenProof: Upon receiving the challenge $chal$, the server
generates a response to prove the data storage correctness.
Specifically, the server chooses a random element $r \leftarrow
\Z_p$, and calculates $R = e(u, v)^r \in \G_T$. Let $\mu'$ denote
the linear combination of sampled blocks specified in $chal$: $\mu'
= \sum_{i \in I} v_i m_i$. To blind $\mu'$ with $r$, the server
computes $\mu = r + \gamma \mu' \mbox{ mod } p$, where $\gamma =
h(R) \in \Z_p$. Meanwhile, the server also calculates an aggregated
authenticator $\sigma = \prod_{i \in I}\sigma_i^{\nu_i}$. It then
sends $(\mu, \sigma, R)$ as the response to the TPA.
\medskip

\noindent VerifyProof: Upon receiving the response $(\mu, \sigma,
R)$ from the cloud server, the TPA validates the response by first
computing $\gamma = h(R)$ and then checking the following
verification equation
\begin{eqnarray}
R\cdot e(\sigma^\gamma,g) \stackrel{?}{=}
e((\prod\limits_{i=s_1}^{s_c}H(W_i)^{\nu_i})^\gamma\cdot u^\mu,v).
\end{eqnarray}
The verification is successful if the equation holds.

\subsection{The Zero Knowledge Public Auditing Protocol}

As pointed out by Wang {\em et al.} in \cite{WangChowWangRenLou},
the TPA protocol presented above is vulnerable to offline guessing
attack. In order to prevent the attack, Wang {\em et al.} proposed
the Zero Knowledge Public Auditing (ZKPA) protocol which is an
extension of their TPA protocol.

The Setup phase is almost the same as in the TPA protocol, except
that an additional generator $g_1 \in \G_1$ is introduced in the
user public key. In the Audit phase, upon receiving the challenge
$chal = \{(i,\nu_i)\}_{i\in I}$, the cloud server selects three
random blind elements $r_m, r_\sigma, \rho \in \Z_p$, and
calculates $R = e(g_1,g)^{r_\sigma} e(u,v)^{r_m} \in \G_T$ and
$\gamma = h(R) \in \Z_p$. The cloud server then calculates $\mu',
\sigma$ according to the TPA protocol (Fig.~\ref{fig:wang}),
blinds both $\mu'$ and $\sigma$ by computing $\mu = r_m + \gamma
\mu' \mod p$, $\varsigma = r_\sigma + \gamma \rho \mod p$ and
$\Sigma = \sigma g_1^\rho$. The cloud server then sends
$(\varsigma, \mu, \Sigma, R)$ as the response to the TPA. To
verify it, the TPA computes $\gamma = h(R)$ and then checks
\begin{eqnarray}
R \cdot e(\Sigma^\gamma,g) \stackrel{?}{=}
e((\prod\limits_{i=s_1}^{s_c}H(W_i)^{\nu_i})^\gamma\cdot
u^\mu,v)\cdot e(g_1,g)^\varsigma.
\end{eqnarray}

\section{Security Weaknesses in Wang \emph{et al.}'s TPA and ZKPA Protocols}\label{sec:attack}

\subsection{Storage Correctness}

It is originally believed that Wang \emph{et al.} TPA protocol can
achieve all the five design goals given in
Sec.~\ref{sec:threatmodel}. However, below we show that the
protocol cannot achieve the important goal of Storage Correctness:
an attacker can arbitrarily modify the data but at the same time
fool the auditor to believe that the data are well maintained by
the cloud server. We describe the attack in two different
scenarios: in the first scenario, the attacker (e.g. a hacker or
internal employee of the cloud server) can learn the content of
the user data file $F$ and modify it; while in the second 
scenario, the attacker can modify the file $F$ but does not know
its content (e.g. a malicious programmer plants a bug in the
software running on the cloud server).

\smallskip\noindent\textbf{Scenario 1:} In this scenario, we assume
the attacker (e.g. an employee of the cloud server) can access the
user data file $F$. The attacker first makes a copy of the original
file, and then modifies file blocks $m_i$ to ${m_i}^* = m_i +
\beta_i$ for $1 \le i \le n$.

In the audit phase, after verifying the file tag $t =
name\|SSig_{ssk}(name)$, the TPA sends a challenge $\{(i,
\nu_i)\}_{i\in I}$ to the cloud server. Upon receiving the
challenge, the cloud server would honestly compute $R = e(u,v)^r$
for a randomly chosen $r$ and $\sigma = \prod_{i \in
	I}\sigma_i^{\nu_i}$. However, as the data file has been modified,
the cloud server would calculate
\begin{eqnarray*}
	\mu^* &=& r + \gamma {\mu'^*} = r + \gamma \sum_{i=s_1}^{s_c} \nu_im_i^*\\
	&=& r + \gamma \sum_{i=s_1}^{s_c} \nu_i(m_i + \beta_i) \\
	&=& \mu + \gamma \sum_{i=s_1}^{s_c} \nu_i \beta_i.
\end{eqnarray*}

When the cloud server sends the response $(\mu^*, \sigma, R)$ to the
TPA, the attacker intercepts the message and generates a new
response as follows:
\begin{enumerate}
	\item compute $\gamma = h(R)$, $\alpha = \gamma \sum\limits_{i=s_1}^{s_c} \nu_i \beta_i$, $\hat R = R \cdot e(u^\alpha, v)$ and $\hat{\gamma} = h(\hat R)$;
	\item compute $\mu' = \sum\limits_{i=s_1}^{s_c}\nu_i m_i$ and $\hat \mu = {\mu'}(\hat \gamma - \gamma) + \mu^*.$
\end{enumerate}
The attacker then sends $(\hat \mu, \sigma, \hat R)$ to the TPA who
will perform the verification according to Equation~(1). The
verification will be successful as shown below.

\begin{eqnarray*}
	\hat R \cdot e(\sigma^{\hat \gamma}, g) &=&  e(u,v)^r e(u^\alpha, v)e((\prod_{i=s_1}^{s_c}\sigma_i^{\nu_i})^{\hat \gamma}, g)\\\\
	& = &e(u^r,v) e(u^\alpha, v)
	e((\prod_{i=s_1}^{s_c} (H(W_i)u^{m_i})^{x\nu_i})^{\hat \gamma}, g)\\
	&=&  e(u^r,v) e(u^\alpha, v)
	e(\prod_{i=s_1}^{s_c}(H(W_i)u^{m_i})^{\nu_i \hat \gamma}, g^x)\\
	&=& e(u^r,v) e(u^\alpha, v)
	e(\prod_{i=s_1}^{s_c} H(W_i)^{\nu_i \hat \gamma}u^{m_i\nu_i \hat\gamma}, v)\\
	&=& e(u^r,v) e(u^\alpha, v)
	e((\prod_{i=s_1}^{s_c} H(W_i)^{\nu_i})^{\hat \gamma} u^{\hat\gamma\mu'}, v)\\
	&=& e((\prod_{i=s_1}^{s_c} H(W_i)^{\nu_i})^{\hat \gamma} u^{\hat\gamma\mu'+\alpha+r}, v)\\
	&=& e((\prod_{i=s_1}^{s_c} H(W_i)^{\nu_i})^{\hat \gamma} u^{\hat\gamma\mu'+ \mu^* - \gamma \mu'}, v)\\
	&=& e((\prod_{i=s_1}^{s_c} H(W_i)^{\nu_i})^{\hat \gamma}
	u^{\hat\mu}, v).
\end{eqnarray*}

\smallskip\noindent\textbf{Scenario 2:} In the second scenario, we assume
the attacker (e.g. a malicious programmer who has planted a software
bug on the cloud server) modifies the file block $m_i$ to ${m_i}^* =
m_i + \beta_i$ for $1 \le i \le n$. However, the attacker only knows
$\beta_i$ (i.e. how the user data are modified) but not $m_i$ or
$m_i^*$.

In the audit phase, the TPA and the cloud server honestly execute
the auditing protocol. That is, TPA sends a challenge $\{(i,
\nu_i)\}_{i\in I}$ to the cloud server, and the cloud server sends
back a response $(\mu^*, R,\sigma)$ where $R = e(u,v)^r$ for a
randomly chosen $r$, $\sigma = \prod_{i \in I}\sigma_i^{\nu_i}$, and
\[
\mu^* = r + \gamma {\mu'^*} = r + \gamma \sum_{i=s_1}^{s_c}
\nu_im_i^*
= r + \gamma \sum_{i=s_1}^{s_c} \nu_i(m_i + \beta_i)
\]
\[
= \mu + \gamma \sum_{i=s_1}^{s_c} \nu_i \beta_i.~~~~~~~~~~~~~~~~~~~~~~~~~~~~~~~~~~~~~~~
\]

The attacker intercepts the response $(\mu^*, R,\sigma)$ from the
cloud server to the TPA, and modifies $\mu^*$  to  $\mu=
\mu^*-\alpha$ where $\alpha = \gamma \sum\limits_{i=s_1}^{s_c}
\nu_i \beta_i = h(R) \sum\limits_{i=s_1}^{s_c} \nu_i \beta_i $. It
is easy to see that by doing such a simple modification, the
attacker derives a correct response with respect to the original
message blocks $\{m_i\}_{i \in I}$. In this way, the attacker can
successfully fool the auditor to believe that the data file $F$ is
well preserved, while the real file on the cloud server has been
modified.

\subsection{Offline Guessing Attack}\label{sec:zk}

The TPA protocol presented in Fig.~\ref{fig:wang} is vulnerable to
offline guessing attack \cite{WangChowWangRenLou}, since the TPA
can always guess whether $\mu' \stackrel{?}{=} \tilde{\mu}'$, by
checking
\begin{eqnarray}
e(\sigma,g) & \stackrel{?}{=} & e((\prod_{i=s_1}^{s_c}
H(W_i)^{\nu_i})\cdot u^{\tilde{\mu}'}, v)
\end{eqnarray}
where $\tilde{\mu}'$ is constructed from random coefficients
chosen by the TPA in the challenge and the guessed message
$\{\tilde{m}_i\}_{s_1 \le i \le s_c}$.

In order to prevent the offline guessing attack, in the ZKPA
protocol, two additional blind elements $r_\sigma$ and $\rho$ are
introduced. It was believed that the ZKPA protocol can effectively
prevent the offline guessing attack. However, below we show that
the ZKPA protocol is still vulnerable to offline guessing attack.
Given $chal = \{(i,\nu_i)\}_{i\in I}$ and the response
$(\varsigma, \mu, \Sigma, R)$, our attack works as follows:
\begin{enumerate}
	\item for the guessed message $\{\tilde{m}_i\}_{s_1 \le i \le
		s_c}$, compute $\tilde{\mu}' = \Sigma_{i\in I} \nu_i \tilde{m}_i$
	and $\tilde{r}_m = \mu - \gamma\tilde{\mu}' \mod p$;
	
	\item compute $e(g_1,g)^{\tilde{r}_\sigma} =
	R/e(u,v)^{\tilde{r}_m}$;
	
	\item compute $e(g_1,g)^{\tilde{\rho}} =
	(e(g,g)^\varsigma/e(g_1,g)^{\tilde{r}_\sigma})^{\gamma^{-1}}$ and
	$e(\tilde{\sigma}, g) = e(\Sigma,g)/e(g_1,g)^{\tilde{\rho}}$;
	
	\item check the equation
	\[e(\tilde{\sigma},g) \stackrel{?}{=} e((\prod_{i=s_1}^{s_c} H(W_i)^{\nu_i})\cdot u^{\tilde{\mu}'}, v).\]
	If the equation holds, then output the guessed message
	$\{\tilde{m}_i\}_{s_1 \le i \le s_c}$; otherwise, go to step (1)
	for another guess.
\end{enumerate}

The above attack essentially shows that the attacker can
successfully remove the additional blind elements introduced in
the ZKPA protocol and use equation (3) to locate the message
$\{m_i\}_{s_1 \le i \le s_c}$.

\section{Conclusion}\label{sec:concl}
In this paper, we revisited a privacy-preserving third party
auditing (TPA) cloud storage integrity checking protocol and its
extended version for zero knowledge public auditing (ZKPA). We
showed several security weaknesses in these protocols. It is still an open problem to design a ZKPA protocol that can prevent offline guessing attacks, and we leave it as our future work.


\end{document}